\documentclass[acus]{JAC2000}
\usepackage{graphicx}

\begin{document}
\title{\flushright{PSN: 3271}\\[15pt] \centering THE NEW MAGNETIC MEASUREMENT 
SYSTEM AT THE ADVANCED PHOTON SOURCE \thanks{Supported by the U.S.Dept. of 
Energy, BES, Office of Science, under Contract No. W-31-109-Eng-38.}}

\author{Yu.Eidelman, B.Deriy, O.Makarov, I.Vasserman, APS, Argonne, IL 60439, USA}

\maketitle

\begin{abstract}
A new system for precise measurements of the field integrals and multipole 
components of the APS
magnetic insertion devices is described. A stretched coil is used to measure
magnetic field characteristics. The hardware includes a number of servomotors 
to move
(translate or rotate) the coil and a fast data acquisition board to measure the
coil signal. A PC under Linux is used as a control workstation. The user 
interface 
is written as a Tcl/tk script; the hardware is accessed from the script  
through a shared C-library. A description of the hardware system and the control 
program is given.
\end{abstract}

\vspace {-.25 cm}
\section{INTRODUCTION}

A new system for precise measurements of magnetic field integrals and multipole 
components of different types of insertion devices (IDs) is created as an 
upgrade of the Magnetic Measurement Facility at the APS \cite {tb_12}. 
The main reasons for upgrading the existing system were:
\begin {itemize}
\vspace {-.2 cm}
\item To increase accuracy and reproducibility of acquired data. 
Long cables between motor drives and stepper motors in the existing 
system produce an excessive noise interfering with the measured signal. 
Integrators in the existing system take 4 data values per coil turn. 
The new system uses $\approx 1000$ data points of coil signal for analysis, leading
to better accuracy. Different error sources
presented in the coil signal can be analyzed, particularly coil 
vibrations.
\vspace {-.2 cm}
\item 
To simplify the system and increase reliability for ease of maintenance,
future developments, and possible duplication of the system in other 
projects. In the existing system VME, Euro, as well as non-standard 
crates, have been used. It has been very hard to maintain and almost 
impossible to reproduce the system.  
\vspace {-.5 cm}
\item To allow integration of the results of the magnetic 
measurements with other APS databases and utilities. The Linux OS instead of 
Windows has to be used in order to reuse software modules developed by 
Operations Analysis Group \cite {oag} for the APS storage ring.
\vspace {-.5 cm}
\item To provide more convenient data analysis in both stand-alone and 
network operation.
\vspace {-.2 cm}
\item To add more features, such as a translation mode, for measuring multipole 
magnetic field components.
\end {itemize}

\vspace {-.5 cm}
\section{Principle of the measurements}

The stretched coil is used to measure magnetic field integrals. The coil
signal is proportional to the time derivative of the magnetic flux 
through the coil crosssection. The measurements are performed 
at constant  coil rotation (translation) speed. There are four different 
options for measurements. 

{\bf Integrals} and {\bf rotation} modes. Signals are measured for full turns 
of the coil. Two turns [clockwise (CW) and counter-clockwise (CCW)] are 
performed to exclude systematic errors. The measured data are applied to 
calculate absolute values of the first or the second (if the coil is twisted 
by 180$^\circ$ \cite {second}) magnetic field integrals. A set of 
approximately ten points 
across the gap of the ID is measured to calculate multipole magnetic moments. 

{\bf Translation} mode. The linear motion of the coil to measure the magnetic 
flux changing in the horizontal or vertical direction is examinated, allowing 
significant reduction in the  measurement time of the multipole moments. 

{\bf Variable field} mode. The coil does not move. This mode is designed for
measurements of IDs with switching electromagnets~\cite {emw}.

\vspace {-.25 cm}
\section{HARDWARE DESCRIPTION}

\begin{figure}[bh]
\centering
\includegraphics*[width=7.5 cm]{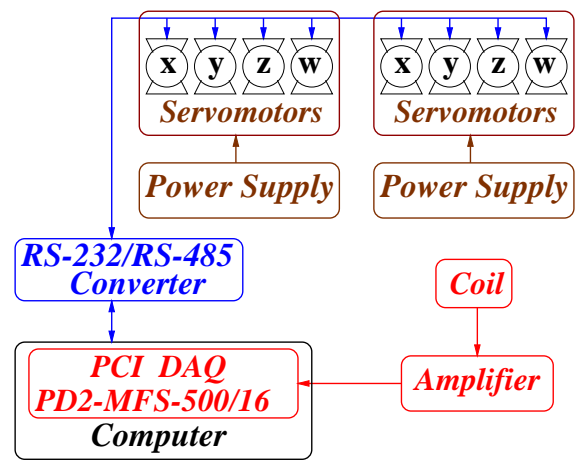}
\caption{Layout of the upgraded system.} 
\label{layout}
\end{figure}

A block diagram of the new system is shown in Fig.~\ref {layout}. 
Eight smart motors from Animatics Corp. \cite {animatics} are used. Each 
servomotor has an  encoder, a motor drive, and a motor controller---all 
integrated in one unit, thus reducing the level of the electrical noise, compared to the 
existing system. These smart motors have two standard serial ports---RS-232 
and RS-485. All motors 
connected in parallel to the RS-485 communication line are controlled by 
the Pentium III computer under Linux OS through a RS-232/RS-485 converter. 

\noindent The stretched coil conventional configuration with parallel wires is
being used for measurements of the first field integrals and multipole
components. The measured signal is amplified by a SCXI-1120 National Instruments 
amplifier. The gain is set to 2000, and the bandwidth is set to 
10~kHz. A 500 kHz data acquisition board PD2-MFS-500/16 from United 
Electronics Industries \cite {uei} is used to measure the amplified signal. 
The Linux driver for this PCI DAQ board was provided by the manufacturer. 

\section{SOFTWARE DESCRIPTION}

A new Multipoles \& Integrals---Software System (MISS) has been written. 
It consists of two parts: a Tcl/tk script 
and a shared C-library. 

\vspace {-.25 cm}
\subsection {Tcl/tk script}

\begin{figure}[h]
\centering
\includegraphics*[width=8.25 cm]{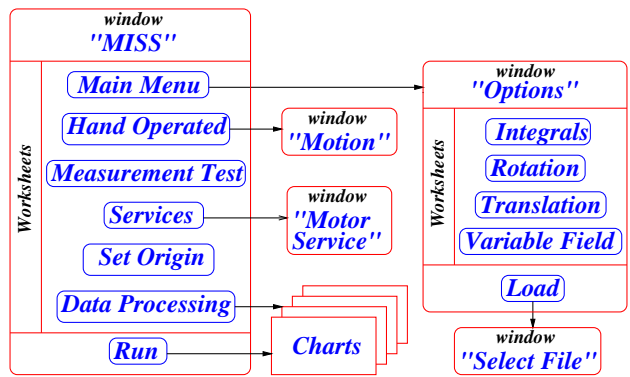}
\caption{Software chartflow diagram.} 
\label{chartflow}
\end{figure}

The Tcl/tk script describes a user interface and data processing and uses a 
shared C-library for hardware access. The script language includes all
embedded instruments to create the user-friendly interface and to
use a shared C-libraries. 
Besides, MISS uses utilities developed by the APS Computer Support Group for 
this language. MISS operates under Linux and provides the 
user interface through the set 
of windows and worksheets (see Fig.~\ref {chartflow}). 

\begin{figure}[tpbh]
\includegraphics[width=8.25 cm]{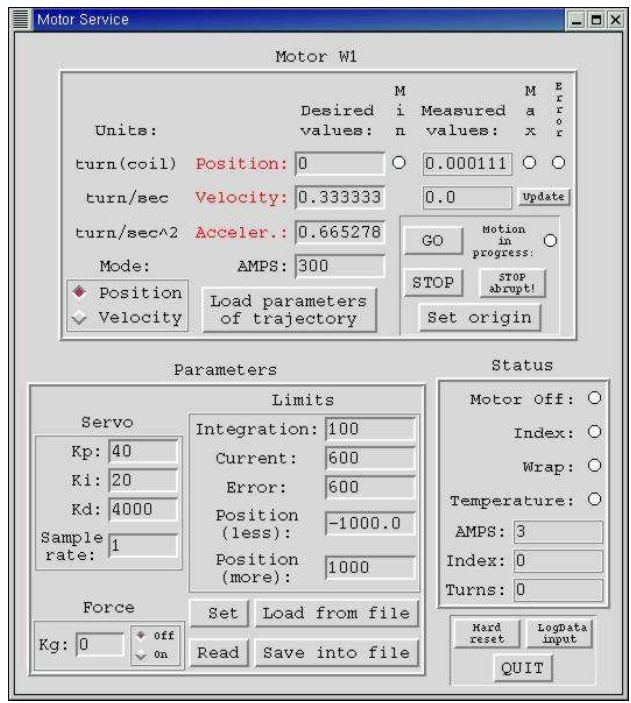}
\caption{Window "Motor Service."} 
\label{motion}
\end{figure}

\begin{figure*}[t]
\centering
\includegraphics*[width=15 cm] {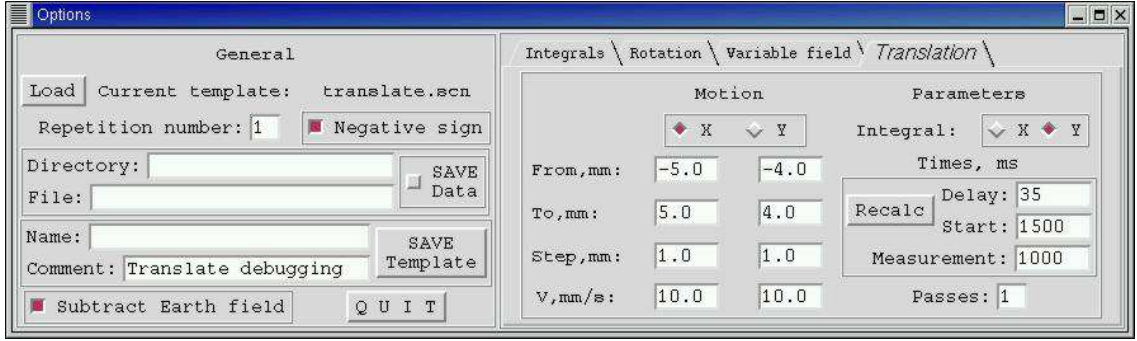}
\caption{Worksheet "Translation" of the window "Options".} 
\label{integrals}
\end{figure*}

\noindent
Figures~\ref {motion}~and~\ref {integrals} demonstrate examples of the user interface.

\vspace {-.25 cm}
\subsection{Shared C-library}

The hardware is accessed from the script through a shared C-library. This 
library consists of two command sets---one for smart motors and another for 
the DAQ board. The serial port driver "ttyS" with
appropriate setting of the structure "termios" is used to
control the motors. 
The Linux driver "pwrdaq" is used to control the DAQ board. Some of the DAQ board 
commands are based on the corresponding C-functions 
of this driver. The complex procedures providing synchronization between 
coil motion and data acquisition are included in the C-library. These procedures
combine calls to the motors and the DAQ board in the required sequence. 
So, the execution of such commands cannot be interrupted by a Tcl/tk script command.

\vspace {-.25 cm}
\section{Synchronization of the coil motion and data acquisition}

This process includes two steps. 

\begin{figure} [htbp]
\includegraphics[width=4 cm] {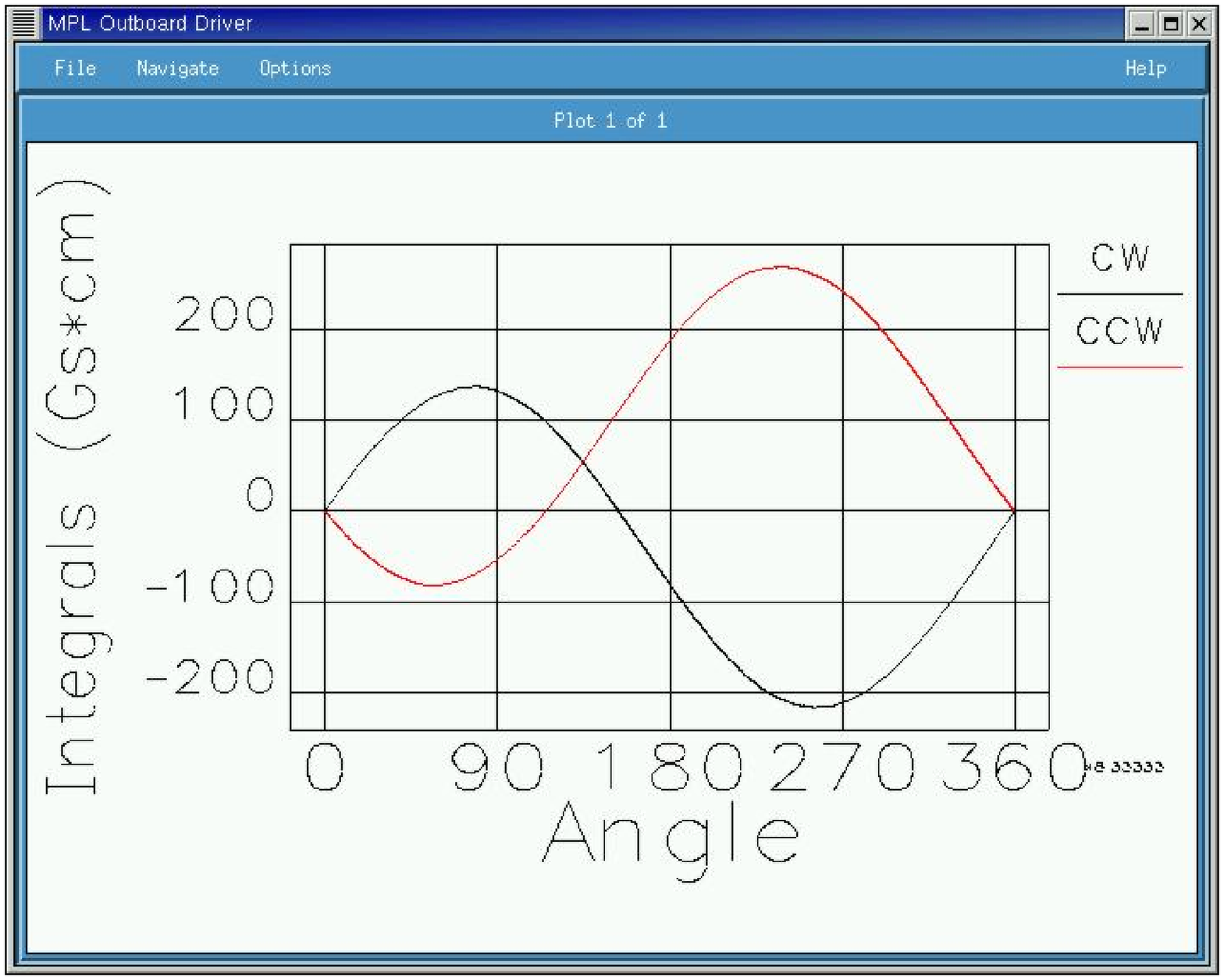}
\includegraphics[width=4 cm] {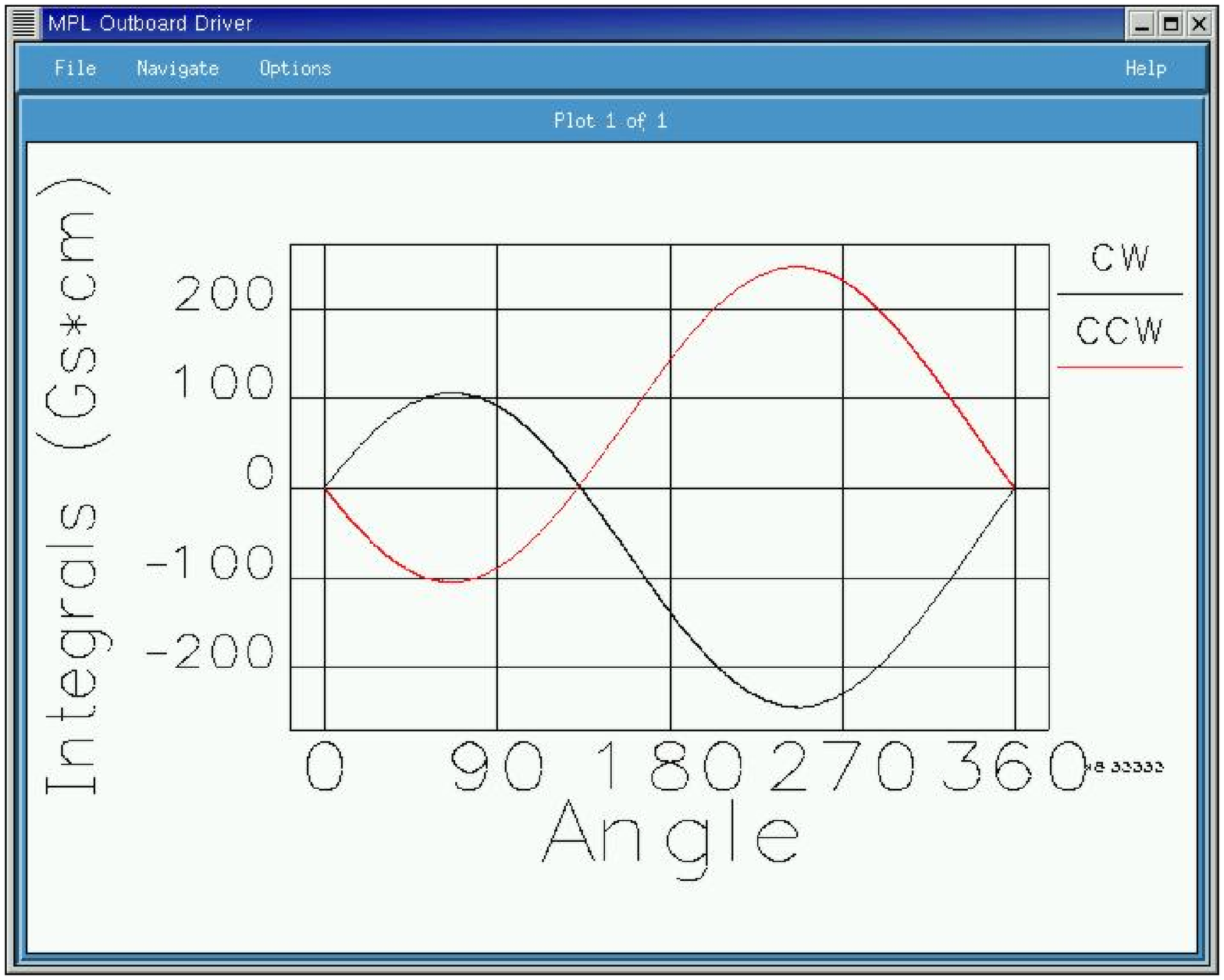}
\caption{Integrals of measured signals for calculated~(left) and matched~(right) 
starts of the data acquisition.} 
\label{synchronization}
\end{figure}

\noindent
{\bf Integrals} and {\bf rotation} modes. At the first step, 
the choice of coil "trajectory" 
(velocity and acceleration of the rotation) determines the calculated
delay between initiating the coil motion and starting the data acquisition. 
The second step includes matching the additional delay to achieve agreement 
between the signals measured for CW  and CCW coil turns. This 
matching is based on the assumption that plots of measured signals (voltage 
versus the coil's turn angle $\alpha$) for both of these rotations 
must be mirror symmetric, if delays are chosen correctly. Figure~\ref 
{synchronization} shows the measured signals for both cases after integration
and subtraction of offsets. It was found that the additional delay is equal to 
25~ms for a calculated delay of 550~ms.

\noindent
{\bf Translation} mode. The first step is identical to the first step in the
rotation mode. The additional delay was determined by using the rotation 
mode with the same velocity and acceleration of the smart motors. This 
delay is equal to 35~ms for a calculated delay of 1500~ms.

\vspace {-.15 cm}
\section {Data processing and some results}

Different algorithms are used to obtain the magnetic properties of the IDs.

\noindent 
A) \underline {Measuring of the first and second integrals}  includes the 
following steps:
\begin {enumerate}
\vspace {-.1 cm}
\item Fitting of the functions $A_i \sin (\alpha + \phi_i)$ on noisy coil 
signals for CW ($i=1$) and CCW coil ($i=2$) turns. 
\vspace {-.2 cm}
\item Calculating of magnetic field integrals $I_x, I_y$ for each coil turn
from fitting parameters $A_i, \phi_i$.
\vspace {-.2 cm}
\item Repetition of previous steps to improve accuracy and define rms errors of
measurements.
%
%
\end {enumerate}

\vspace {-.2 cm}
\noindent
The statistical error of integrals  $I_x, I_y$ (measured for the Earth's magnetic
field) is equal to $\pm~0.4$~Gs$\cdot$cm (improved from $\pm~2$~Gs$\cdot$cm 
before upgrading).

\vspace {.25 cm}
\noindent 
B) \underline {Measuring of the integrated multipole moments} in the rotation
mode includes all steps from case A) for a set of points $r_i$ across the ID gap 
and, after that:
\begin {enumerate}
\setcounter {enumi} {3}
\vspace {-.1 cm}
\item Fitting of the polynomials $\sum_{k=0}^n b_k r^k$ of the power $n$ on
 the sets $I(r)$ (separately for arrays $I_x$ and $I_y$).
\end {enumerate}

\noindent 
Coefficients $b_k$ are integrated multipole moments. 

\vspace {.25 cm}
\noindent 
C) \underline {Measuring of the integrated multipole moments} in the 
translation mode includes:
\begin {enumerate}
\vspace {-.1 cm}
\item Fitting of the polynomials $\sum_{k=0}^n b_k r^k$ of the power $n$ on
noisy signals for a few coil passes across the gap of the ID. 
\vspace {-.2 cm}
\item Averaging of obtained coefficients $b_k$ over the set of coil passes to 
reduce the statistical errors.
\end {enumerate}

\noindent 
Note, that the translation mode saves significant
time in measuring  the integrated multipole moments in comparison with the 
rotation mode. Measurements have to be done for both horizontal and vertical 
coil orientations.

\noindent 
A strong short magnet was used to evaluate the accuracy of the measurements 
of the multipole 
moments. Table~\ref {b_k} shows these results which were obtained 
for the translation mode.
\vspace {-.5 cm}
\begin {table} [htpb] 
\caption {Multipole moments measurements.}
\begin {center}
\begin {tabular} {|c|c|c|c|} \hline
$k$ & $b_k$ & $\delta b_k$ & Units \\ \hline \hline
0 & 652.8 &  $\pm$0.1 & Gs$\cdot$cm \\ \hline
1 & 2208 &  $\pm$2 & Gs \\ \hline
2 & 1840 &  $\pm$7 & Gs$\cdot$cm$^{-1}$  \\ \hline
3 & -1167 & $\pm$48 & Gs$\cdot$cm$^{-2}$ \\ \hline
4 & -2929 & $\pm$127 & Gs$\cdot$cm$^{-3}$ \\ \hline
5 & 1985 & $\pm$203 & Gs$\cdot$cm$^{-4}$  \\ \hline
6 & 460 & $\pm$518 & Gs$\cdot$cm$^{-5}$  \\ \hline
\end {tabular}
\end {center}
\label {b_k} 
\end {table}

\vspace {-.5 cm}
\noindent 
A few programs from the SDDS Toolkit \cite {borland} are used for polynomial 
fitting, data management, and displaying charts.

\end{document}